\documentclass[twocolumn,aps,showpacs]{revtex4-1}
\usepackage{amsfonts}
\usepackage{amsmath}
\usepackage{graphicx}
\usepackage{bm}
\usepackage{amssymb}
\usepackage{dcolumn}
\usepackage{color}

\begin{document}

\title{Enhanced spin Hall effect by electron correlations in CuBi alloys}

\author{Bo Gu$^{1,2}$,                                                                                           
Zhuo Xu$^{1,2}$, Michiyasu Mori$^{1,2}$, Timothy Ziman$^{3,4}$, and Sadamichi Maekawa$^{1,2}$}   

\affiliation{$^1$Advanced Science Research Center, Japan Atomic Energy Agency, Tokai 319-1195, Japan             
\\ $^{2}$CREST, Japan Science and Technology Agency, Sanbancho, Tokyo 102-0075, Japan                            
\\ $^{3}$ Institut Laue Langevin, Bo\^\i te Postale 156, F-38042 Grenoble Cedex 9, France                        
\\ $^{4}$ LPMMC  (UMR 5493), Universit\'e Grenoble 1 and CNRS, 38042 Grenoble, France} 

\begin{abstract}
A recent experiment in CuBi alloys obtained a large spin Hall angle (SHA) of -0.24 (Niimi {\it et al}. Phys. Rev. Lett. {\bf 109}, 156602 (2012)). We find that the SHA can be dramatically enhanced by Bi impurities close to the Cu surface. The mechanisms of this enhancement are two-fold. 
One is that the localized impurity state on surface has a decreased hybridization and 
 combined with  Coulomb correlation effect. 
The other comes from the low-dimensional state of conduction electrons on surface, which results in a further enhancement of skew scattering by impurities. 
Furthermore, we note that a discrepancy in sign of SHA between the experiment and previous theories is simply caused by different definitions of SHA. This re-establishes skew scattering as the essential mechanism underlying the spin Hall effect in CuBi alloys.
\end{abstract}

\pacs{71.70.Ej, 75.30.Kz, 75.40.Mg} 
\maketitle

\section{Introduction}
Spin Hall effect (SHE), by which an input charge
current is converted into a transverse spin current, is one of important phenomena in spintronics~\cite{Dyakonov, SHE-Hirsch}. 
For practical applications, it is necessary to enhance the ratio of the induced spin current to the input charge current, i.e., spin Hall angle (SHA).  
The spin-orbit interaction (SOI) is the origin of the SHE. 
So far, however, even in heavy elements such as Pt, the SHA was about several percent at most. 
Recently, Niimi {\it et al.} has reported that Cu including a few percent of Bi shows a large SHA of -0.24 at low temperatures~\cite{CuBi-Niimi}.  
Apparently, 24 \% of SHA is one order larger than that of Pt. 
Hence, the mechanism behind their surprising result becomes the interesting subject of condensed matter physics and simultaneously the important issue of spintronics.    

A possible mechanism is the skew scattering of $6p$ orbitals of Bi, i.e., spin-dependent deflection of the scattered electrons due to the spin-orbit interaction of the Bi impurities in Cu~\cite{CuBi-Niimi, CuBi-Gradhand}. 
Gradhand {\it et al.}  predict  the {\it positive} SHA of 0.081 using the density functional theory (DFT) and the semi-classical Boltzmann equation~\cite{CuBi-Gradhand}.  
On the other hand, Niimi et al. {\cite{CuBi-Niimi}} estimate the SHA induced by the resonant skew scattering of $6p$ orbitals on Bi impurities~\cite{Fert-Levy}, and report 
the negative SHA of -0.046 by using a phase shift obtained in the DFT calculation with the spin-orbit interaction. 
It is discussed that the large negative SHA of -0.24 could be explained by increasing the splitting of phase shift by a factor of three~\cite{CuBi-Niimi}. 
Fedorov {\it et al.} have also presented an extended phase shift model 
(essentially the same as that of Refs.~\cite{CuBi-Niimi} and~\cite{Fert-Levy}), 
and obtained the SHA by $6p$ orbitals of impurities. 
However, their result is still less that half of the experimental result by Niimi in the magnitude and furthermore with opposite sign \cite{CuBi-Fedorov}. 
They conclude that the discrepancy between the experiment and their theory 
can not be explained by the conventional skew-scattering contribution, the side-jump contribution or the intrinsic mechanisms \cite{CuBi-Fedorov}. 
Therefore, the large magnitude of SHA in the CuBi alloy is still an open question. In addition, the discrepancy between theory and experiment comes out. 


In this paper, we examine the theory of the SHE due to skew scattering of impurities with $6p$ orbitals. 
We find that the SHA can be dramatically enhanced  by the presence of  Bi impurities on the Cu surface combined with the electrons correlation effect. 
In addition, we find that the apparent discrepancy in sign of SHA in CuBi alloys came  from a confusion of definitions: 
the SHA, if  defined as a ratio between transverse and longitudinal resistances, $\alpha(\rho) =\rho_{yx}/\rho_{xx}$,  is of  opposite sign to  the  ratio between transverse and longitudinal {\it conductivities, $\alpha(\sigma)=\sigma_{yx}/\sigma_{xx}$}.
Once the confusion of definitions is removed, there is no contradiction to a skew scattering mechanism, and thus we restore the possibility of properly microscopic understanding of the effects.

\section{Skew scattering by impurities of $p$ orbital in bulk}

We now derive the expression in the 
formalism that has been applied to the SHE in n-doped GaAs \cite{Engel} ,
as well as in Au metal with impurities of $d$ orbitals \cite{Guo, Gu-AuFe, Gu-AuPt}.  
We consider the scattering of conduction electrons by non-magnetic impurities of $p$ orbitals with spin-orbit interaction. 
The Hamiltonian of the system is described in Ref.\cite{Bethe-Jackiw}.
Taking the incoming wave as a three-dimensional (3D) plane wave, the amplitude of the scattered wave is given by \cite{Bethe-Jackiw}
\begin{equation}
\begin{split}
f_{\uparrow}(\theta) &= f_1(\theta) |\uparrow\rangle + e^{i\varphi}f_2(\theta)|\downarrow\rangle, \\
f_{\downarrow}(\theta) &= f_1(\theta) |\downarrow\rangle - e^{-i\varphi}f_2(\theta)|\uparrow\rangle,
\end{split}
\label{E-f-up-dn-3d}
\end{equation}  
for incoming spin-up and spin-down electrons, respectively. $\theta$ and $\varphi$  are the  polar and azimuthal angles of the scattered wave vector $\bf{k}^{\prime}$ in  coordinates  where the incident wave vector $\bf{k}$ is  along $z$ axis. $f_1(\theta)$ and $f_2(\theta)$ describe the spin-conserving and spin-flip scattering amplitudes. For incoming electron with general spin state $|\xi\rangle$ = $a|\uparrow\rangle$ + $b|\downarrow\rangle$, 
the differential scattering cross section is given by 
$\frac{d\sigma}{d\Omega}$ = $I(\theta)$ + $I(\theta)S(\theta)\pmb{\sigma}\cdot\pmb{n}$ \cite{Bethe-Jackiw}, 
where $\pmb{\sigma}$ is a vector of Pauli matrix, $\pmb{n}$ = $\pmb{k}\times\pmb{k}^{\prime}/|\pmb{k}\times\pmb{k}^{\prime}|$, 
and $I(\theta)$ and $S(\theta)$ are the spin-independent part and skewness of the scattering cross section,
\begin{equation}
I(\theta) = |f_1(\theta)|^2 + |f_2(\theta)|^2, 
S(\theta) = \frac{2Im\left[f^{\ast}_1(\theta)f_2(\theta)\right]}{|f_1(\theta)|^2 + |f_2(\theta)|^2}. 
\label{E-I-S-3d}
\end{equation}
In terms of phase shift $\delta$ of partial wave $|\ell,m\rangle$, the amplitudes are \cite{Bethe-Jackiw} 
\begin{equation}
\begin{split}
f_1(\theta) &= \sum_{\ell,m}\frac{P_{\ell}(\cos\theta)}{2ik}[(\ell+1)(e^{2i\delta_{\ell+\frac{1}{2}}}-1) 
+\ell (e^{2i\delta_{\ell-\frac{1}{2}}}-1)], \\
f_2(\theta) &= \sum_{\ell,m}\frac{-\sin(\theta)}{2ik}\left[e^{2i\delta_{\ell+\frac{1}{2}}}-e^{2i\delta_{\ell-\frac{1}{2}}}\right]
\frac{d}{d\cos\theta}P_{\ell}(\cos\theta).
\end{split}
\label{E-f1-f2-3d}
\end{equation}
Following the derivation of Ref. \cite{Engel}, 
the distribution function $f_{\pmb{k}}$ can be obtained as
$f_{\pmb{k}} = f^0_k + \pmb{k} \cdot \left[\pmb{E}
+\frac{\gamma_k}{2}\left(\pmb{\sigma}\times\pmb{E}\right)\right]C_k$, 
where $f^0_k$ is the equilibrium distribution function, $k = |\pmb{k}|$, $\pmb{k}$ is crystal momentum, $\pmb{E}$ is external electric field,
and $C_k$ is a coefficient. Note that the factor $1/2$ arises from the integration on the azimuthal angle $\varphi$ in a three-dimensional system. 
The  transport skewness $\gamma_k$ is defined as \cite{Engel}
\begin{equation}
\gamma_k=\frac{\int d\Omega I(\theta)S(\theta)\sin\theta}{\int d\Omega I(\theta)(1-\cos\theta)}
\label{E-skewness}
\end{equation}
where $\Omega$ is the solid angle in three dimensions; the  polar angle $\theta$ is defined as the angle between $\pmb{k}$ and $\pmb{k}^{\prime}$.
The currents $\pmb{j}^{(\pm)}$ can be expressed as 
$\pmb{j}^{(\pm)} = \left<\pm\left|\int\frac{d^3\pmb{k}}{(2\pi)^3}\frac{\hbar\pmb{k}}{m}f_{\pmb{k}}\right|\pm\right>$,
where $\pm$ denote the  spins  of the conduction electrons. 
Considering the case $\pmb{\sigma}=(0,0,\sigma_z)$ and $\pmb{E} = (E_x,0,0)$, we obtain
$\pmb{j}^c \equiv \pmb{j}^{(+)} + \pmb{j}^{(-)} = 
\hat{\pmb{x}}\left[\int\frac{d^3\pmb{k}}{(2\pi)^3}\frac{\hbar k^2}{3m}C_k\right]2E_x$ 
and
$\pmb{j}^s \equiv \pmb{j}^{(+)} - \pmb{j}^{(-)} = 
\hat{\pmb{y}}\left[ \left(\frac{\gamma_{k_F}}{2}\right)\int\frac{d^3\pmb{k}}{(2\pi)^3}\frac{\hbar k^2}{3m}C_k\right]2E_x$.
The above two equations can be written in terms of electric conductivities
$j^{c}_x$ = $j^{(+)}_x$ +$j^{(-)}_x$ = $(\sigma^{(+)}_{xx}$ +$\sigma^{(-)}_{xx})E_x$ = $2\sigma^{(+)}_{xx}E_x$ and 
$j^{s}_y$ = $j^{(+)}_y$ -$j^{(-)}_y$ = $(\sigma^{(+)}_{yx}$ -$\sigma^{(-)}_{yx})E_x$ = $2\sigma^{(+)}_{yx}E_x$.
The spin Hall angle $\alpha (\sigma)$ can be defined by currents 
$\pmb{j}^{s} \equiv \alpha(\sigma)\left(\hat{\pmb{z}}\times \pmb{j}^{c}\right)$ \cite{Maekawa}.
It can then be related to the transport skewness of equation (\ref{E-skewness}),
$\alpha(\sigma) = j^s_y/j^c_x=\sigma^{(+)}_{yx}/\sigma^{(+)}_{xx}=\gamma_{k_F}/2$. 
By Eqs. (\ref{E-I-S-3d}) and (\ref{E-f1-f2-3d}), we finally obtain
\begin{equation}
\begin{split}
&\alpha (\sigma)  = \\
&\frac{2\sin\delta_0\left[\sin(\delta_{1/2}-\delta_0)\sin\delta_{1/2}-\sin(\delta_{3/2}-\delta_0)\sin\delta_{3/2}\right]}
{3\left(\sin^2\delta_0+\sin^2\delta_{1/2}+2\sin^2\delta_{3/2}\right)}.
\end{split}
\label{E-alpha-sigma-phase}
\end{equation}
This recovers the expression of Fedorov {\it et al.} \cite{CuBi-Fedorov}.
If instead of defining a SHA as a ratio of currents, we do so as that of resistances, as is common in the experimental literature,
then from  the general relation 
$\rho_{yx}$ = - $\sigma_{yx}/\sigma^2_{xx}$,
we have 
$\alpha(\rho ) \equiv \rho^{(+)}_{yx}/\rho^{(+)}_{xx}= -\alpha (\sigma)$.
The apparent discrepancy in sign of SHA in CuBi alloys between the experiment \cite{CuBi-Niimi} and previous theories \cite{CuBi-Gradhand, CuBi-Fedorov} can thus be eliminated by taking care over the definitions used.

\section{Skew scattering by impurities of $p$ orbital on a surface}
We consider the scattering of conduction electrons by impurities of $p$ orbitals with spin-orbit interactions,  within a very simplified approach:
taking a two-dimensional (2D) incoming plane wave on the surface
 defined as the $x$-$y$ plane \cite{2d-pw-sax,2d-pw-zeng} , 
 and considering the outgoing scattered wave projected into the same plane using the (3D) scattering amplitudes.   
Similarly to the calculation leading to Eq. (\ref{E-f-up-dn-3d}), for a 2D plane wave we obtain the following amplitude of the scattered wave 
\begin{equation}
\begin{split}
f_{\uparrow}(\varphi) = f_{1\uparrow}(\varphi) |\uparrow\rangle, \quad
f_{\downarrow}(\varphi) = f_{1\downarrow}(\varphi) |\downarrow\rangle,
\end{split}
\label{E-f-up-dn-2d}
\end{equation} 
for incoming spin-up and spin-down electrons, respectively. $\varphi$ is azimuth angle of scattered wave vector $\bf{k}^{\prime}$ in the coordinate where the incident wave vector $\bf{k}$ is set along $x$ axis, where polar angle $\theta$ = $\pi/2$ for $\pmb{k}$ and $\pmb{k}^{\prime}$. 
In contrast to the  bulk case, here there are  only spin-conserving scattering amplitudes $f_{1\uparrow}(\varphi)$ and $f_{1\downarrow}(\varphi)$, and no spin-flip scattering amplitudes. For an incoming electron with general spin state $|\xi\rangle$ = $a|\uparrow\rangle$ + $b|\downarrow\rangle$, 
the differential scattering cross section is given by 
$\frac{d\sigma}{d\Omega}$ = $I(\varphi)$ + $I(\varphi)S(\varphi)\pmb{\sigma}\cdot\pmb{n}$, 
where $I(\varphi)$ and $S(\varphi)$ are the spin-independent part and skewness of the scattering cross section,
\begin{equation}
I(\varphi) = |g(\varphi)|^2 + |h(\varphi)|^2, 
S(\varphi) = \frac{2Re\left[g^{\ast}(\varphi)h(\varphi)\right]}{|g(\varphi)|^2 + |h(\varphi)|^2}. 
\label{E-I-S-2d}
\end{equation}
Here $g(\varphi)$ = $[f_{1\uparrow}(\varphi)+f_{1\downarrow}(\varphi)]/2$, and $h(\varphi)$ = $[f_{1\uparrow}(\varphi)-f_{1\downarrow}(\varphi)]/2$.
In terms of phase shift $\delta$ of the lowest partial waves $m=0$ and $1$ , it has
\begin{equation}
\begin{split}
g(\varphi) &= \frac{e^{-i\pi/4}}{\sqrt{2\pi k}}[(e^{2i\delta_{0}}-1)+\frac{4}{3}\cos{\varphi}(e^{2i\delta_{3/2}}-1) \\
&+\frac{2}{3}\cos{\varphi}(e^{2i\delta_{1/2}}-1)], \\
h(\varphi) &= \frac{e^{-i\pi/4}}{\sqrt{2\pi k}}(\frac{2}{3}i\sin{\varphi})(e^{2i\delta_{3/2}}-e^{2i\delta_{1/2}}).
\end{split}
\label{E-g-h-2d}
\end{equation} Note that the non-zero skewness $S(\varphi)$ originates from the SOI in $p$ orbitals, i.e., $\delta_{3/2} \neq \delta_{1/2}$. 
As observed in Ref. \cite{Engel}, 
the distribution function $f_{\pmb{k}}$ can be obtained as
$f_{\pmb{k}} = f^0_k + \pmb{k} \cdot \left[\pmb{E}+\gamma^{2D}_k\left(\pmb{\sigma}\times\pmb{E}\right)\right]C_k$, 
where the transport skewness $\gamma^{2D}_k$ is now determined by integration over a single angle
\begin{equation}
\gamma^{2D}_k=\frac{\int^{\pi}_0 d\varphi I(\varphi)S(\varphi)\sin\varphi}{\int^{\pi}_0  d\varphi I(\varphi)(1-\cos\varphi)}.
\label{E-skewness-2d}
\end{equation}
Just as  in the bulk  
$\alpha^{2D}(\sigma) = \sigma^{+}_{yx}/\sigma^{+}_{xx}=\gamma^{2D}_{k_F}$.
By Eqs. (\ref{E-I-S-2d}) and (\ref{E-g-h-2d}), we finally obtain
\begin{equation}
\begin{split}
&\alpha^{2D} (\sigma) = \\
&\frac{2\sin\delta_0\left[\sin(\delta_{1/2}-\delta_0)\sin\delta_{1/2}-\sin(\delta_{3/2}-\delta_0)\sin\delta_{3/2}\right]}
{3\sin^2\delta_0+2\left(\sin^2\delta_{1/2}+2\sin^2\delta_{3/2}\right)}.
\end{split}
\label{E-alpha-sigma-phase-2d}
\end{equation}
By the general relation $\rho_{yx}$ = - $\sigma_{yx}/\sigma^2_{xx}$, 
we also have 
$\alpha^{2D}(\rho ) \equiv \rho^{(+)}_{yx}/\rho^{(+)}_{xx}= -\alpha^{2D} (\sigma)$.
While Eq. (\ref{E-skewness-2d}) agrees with the general 2D expression cited in Ref. \cite{Engel}, we emphasize that the 
functions  $I(\varphi)$ and $S(\varphi)$ here come from the resonant scattering, 
as defined in Eqs. (\ref{E-I-S-2d}) and (\ref{E-g-h-2d}) based on a phase shift analysis, 
therefore are quite different from those by the screened Coulomb interactions in Ref. \cite{Engel}. 
As a result, our expression for spin Hall angle for 2D case in Eq. (\ref{E-alpha-sigma-phase-2d}) is novel. 
Comparing Eqs. (\ref{E-alpha-sigma-phase-2d}) and (\ref{E-alpha-sigma-phase}) with same phase shift parameters, 
it is found that $\alpha^{2D}(\rho)/\alpha(\rho)$$\rightarrow$ 1.5 in the limit where $p$ phase shifts $\delta_{3/2}$ and $\delta_{1/2}$ dominate.    
Thus within our simplified approach, the difference of taking purely planar  scattering  leads to an   increase in the observed spin hall angle which can be quite substantial.
\begin{figure}[tbh]
\includegraphics[width = 6.5 cm]{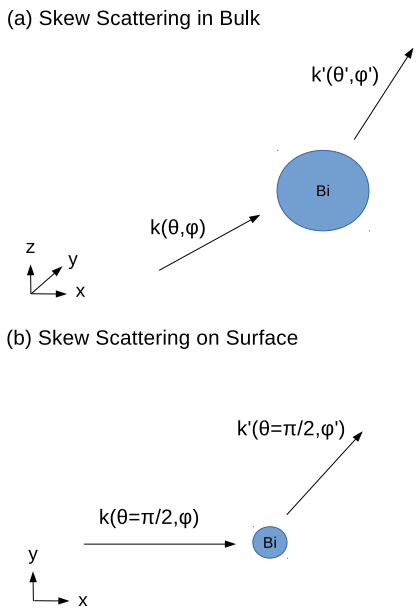}
\caption{Schematic picture of skew scattering of conduction electrons by the Bi impurity in Cu bulk (a) and on a Cu surface (b). The Bi impurity
 has a very extended state (large circle) in the bulk and a much more localized state (small circle) on the surface. 
The conduction electron in Cu is described by a three-dimensional
wave vector $\pmb{k}(\theta,\varphi)$ in the bulk, and a two-dimensional wave vector $\pmb{k}(\theta = \pi/2,\varphi)$ on the surface, 
where $\theta$ and $\varphi$ are polar and azimuthal angles respectively. }
\label{F-schematic}
\end{figure}

\section{Numerical results for $CuBi$}
We turn to the specific case of Bi impurities in Cu, comparing experiment and theory, while   the issue of {\it sign} is now resolved, there remains that of the magnitude.  The authors of Ref. \cite{CuBi-Niimi} found $\alpha(\rho)$ = -0.046  from phase shift parameters $\delta_{1/2}$ = 1.18, $\delta_{3/2}$ = 0.72, and $\delta_{0}$ = 1.46. They  pointed out that only by adjusting by hand from the values estimated by density functional theory (DFT),  would the difference in phase shifts   for the  spin-orbit split p-states  be sufficient to explain experiment, and suggested that the Coulomb correlation effect, following Ref. \cite{Gu-AuFe}, might produce such a renormalization.
This we have tested with Hirsch-Fye Quantum Monte Carlo (QMC) method \cite{QMC}, 
but for simple substitutional impurities the numerical results do not bear out this hope.
In our DFT calculations \cite {QE} with the local density approximation (LDA), we find that the hybridization between Bi impurity and Cu bulk is very large ($\approx$ 15 eV at $\Gamma$ point), which is much larger than the Coulomb correlation $U \approx$ 4 eV for the $6p$ orbitals of Bi impurities. 
Thus it is not surprising that the Coulomb correlation $U$ does not appear to enhance the SHA for a Bi impurity in bulk Cu.
We therefore considered the possibility that surface enhancement of the impurity scattering might play a role.

By LDA+SOI calculation \cite{QE}, we study the SHE due to the skew scattering by a single Bi impurity on Cu (111) surface.  
We consider two different possible surface dopings; substitution of a surface atom and intercalation of the impurity close to, but slightly above, the copper surface.
Our results are listed in Table \ref{T-surface}. In both cases the SHA $\alpha(\rho)$ = -0.077, which increases the magnitude compared to the 
SHA $\alpha(\rho)$ = -0.046 in the bulk and remains of the same sign. The increase is primarily because of the localization of impurity state on surface, as shown in Fig. \ref{F-schematic}.
We find that the hybridization between Bi impurity and Cu surface is dramatically decreased ($\approx$ 4 eV at $\Gamma$ point), compared to the bulk case, and is now  close to the Coulomb correlation $U \approx$ 4 eV for the $6p$ orbitals of Bi impurities. 

\begin{table}
\begin{tabular}{cccccc}
\hline
On Surface       & $\delta_0$ & $\delta_{1/2}$ & $\delta_{3/2}$ & $\alpha(\rho)$ & $\alpha^{2D}(\rho)$ \\
\hline
LDA+SOI (s)      & 1.71    & 1.49  &  0.75 & -0.077& -0.099    \\
LDA+SOI (i)      & 1.79    & 1.58  &  0.58 & -0.077& -0.098    \\
QMC(U=4eV) (i)   & 1.79    & 1.87   & 0.33  & -0.125 & -0.152 \\
\hline
\end{tabular}
\caption{Numerical result for the skew scattering due to a Bi impurity on the Cu surface. Two possible surface dopings: substitution (s) of a surface atom and intercalation (i) of the impurity close to, but slightly above the surface.}
\label{T-surface}
\end{table}

Next, by a combined theoretical approach of LDA+QMC \cite{Gu-AuFe,Gu-AuPt} 
we study the Coulomb correlation effect on the SHE. It is a two-step calculation. 
First, a single-impurity multi-orbital Anderson model \cite {Anderson} is formulated within the LDA, 
for determining the detailed host band structure, the
impurity levels, and the impurity-host hybridization. 
Second, the electron correlations in this Anderson model at finite
temperatures are calculated by the Hirsch-Fye QMC method \cite{QMC}.
The single-impurity multi-orbital Anderson model is defined as
\begin{eqnarray}
  H&=&\sum_{\textbf{k},\alpha,\sigma}\epsilon_{\alpha}(\textbf{k})
  c^{\dag}_{\textbf{k}\alpha\sigma}c_{\textbf{k}\alpha\sigma}
   +\sum_{\textbf{k},\alpha,\xi,\sigma}(V_{\xi\textbf{k}\alpha }
    p^{\dag}_{\xi\sigma} c_{\textbf{k}\alpha\sigma} + H.c.) \notag\\
  &+& \sum_{\xi,\sigma}\epsilon_{\xi}n_{\xi\sigma} 
   +U\sum_{\xi}n_{\xi\uparrow}n_{\xi\downarrow} 
   + \frac{U^{\prime}}{2}\sum_{\xi\neq\xi',\sigma,\sigma^{\prime}}
     n_{\xi\sigma}n_{\xi'\sigma^{\prime}} \notag\\
   &-& \frac{J}{2}\sum_{\xi\neq\xi',\sigma}n_{\xi\sigma}n_{\xi'\sigma}
  +\frac{\lambda}{2}\sum_{\xi,\sigma}p^{\dagger}_{\xi\sigma}
   (\ell)^{z}_{\xi\xi}(\sigma)^{z}_{\sigma\sigma}p_{\xi\sigma},
\label{E-Ham}
\end{eqnarray}
where 
$\epsilon_{\alpha}(\textbf{k})$ is host energy band,
$\epsilon_{\xi}$ is impurity energy levels, 
$V_{\xi\textbf{k}\alpha}$ is impurity-host
hybridization,
$U$ ($U'$) is the on-site Coulomb repulsion within (between) the orbitals of the
impurity, $J$ is the Hund coupling between the orbitals of the impurity.
Considering the relationship $U$ = $U^{\prime}$ + 2$J$ \cite{SM-TT}, in our QMC calculations,
we use the values of $U$ = 4 eV, $U^{\prime}$ = 1.6 eV, and $J$ = 0.8 eV.
The value of the SOI of $6p$ orbitals of Bi atom is $\lambda$ = 1.45 eV \cite{SOI-Bi}. 
For simplicity we consider only the $z$ component of the SOI.

The QMC results are also listed in Table \ref{T-surface}. 
Including the on-site correlation effect for the $6p$ orbitals of Bi impurity 
with $U$ = 4 eV, $U^{\prime}$ = 1.6 eV, and $J$ = 0.8 eV, 
which can be correctly treated by QMC calculations, further increases the splitting and predicts a SHA $\alpha(\rho)$ = -0.125, 
which further increases the magnitude compared to $\alpha(\rho)$ = -0.077 by LDA+SOI. 
Using the formula corresponding to our simplified approach to   two-dimensionality, the SHA becomes  $\alpha^{2D}(\rho)$ = -0.152, 
a  further increase in magnitude. While this final  estimate
of SHA of -0.152 is still smaller  than that of -0.24 recently observed in CuBi alloys \cite{CuBi-Niimi},  the values are similar enough that the skew scattering mechanism must now be considered. 

Niimi {\it et al} argue for a possible 1D-to-3D effect to explain their experimental results in CuBi alloys\cite{CuBi-Niimi, CuBi-Niimi-2}. Our result here suggests another possible mechanism to enhance the spin Hall effect by considering Bi impurities on Cu surface within a simplified approach. A more rigorous treatment of surface may be helpful in future.  

\section{Conclusion}
In summary, we have re-examined  the theory of the SHE in CuBi alloys due to resonant skew scattering by 6$p$ orbitals.
We showed that the apparent discrepancy in sign of the SHE in CuBi alloys between the experiment and the established theories 
can be removed simply by using consistent definitions. This re-establishes skew scattering as the essential mechanism underlying the SHE and  AHE in these materials.
We found that the SHA can be dramatically enhanced if  Bi impurities occur on the Cu surface by two effects.
Firstly, the decreased hybridization of the localized impurity state on the surface, combined with the Coulomb correlation, enhances SHE. Secondly the two-dimensional state of conduction electrons on surface can  result in a further enhancement of skew scattering by impurities. The combined effects may  be responsible for the large SHE recently observed in CuBi alloys \cite{CuBi-Niimi}.

\section*{Acknowledgement}		
The authors acknowledge  Y. Niimi and Y. Otani for many valuable
discussions about the experiments of spin Hall effect in CuBi alloys and A. Fert, P. Levy, I. Mertig and G. Y. Guo for helpful and encouraging correspondence. This work was supported by Grant-in-Aid for Scientific Research from MEXT (Grant No.24540387, No.26108716, No.26247063, and No.26013006), by bilateral program from MEXT, by CREST from JST, by REIMEI from JAEA, and by National Science Foundation under Grant No. NSF PHY11-25915.

\end{document}